# Statistical Analysis and Stochastic Dislocation-Based Modelling of Microplasticity

O. Kapetanou[1], V. Koutsos[1], E. Theotokoglou[2], D. Weygand[3], M. Zaiser[1,4]

[1]School of Engineering, Institute for Materials and Processes, The King's Buildings, Sanderson Building, Edinburgh EH93JL, UK

[2] School of Applied Mathematical and Physical Sciences Department of Mechanics Laboratory of Testing and Materials National Technical University of Athens Zographou Campus, Theocaris Bld GR-157 73, Athens, Greece

[3]Institute for Applied Materials, Karlsruhe Institute of Technology, Kaiserstrasse 12, 76199 Karlsruhe, Germany

[4]Department of Materials Science, Institute for Materials Modelling, University of Erlangen-Nürnberg, Dr.-Mack-Strasse 77, 90762, Fürth, Germany

**Abstract**. Plastic deformation of micro- and nanoscale samples differs from macroscopic plasticity in two respects: (i) the flow stress of small samples depends on their size (ii) the scatter of plastic deformation behaviour increases significantly. In this work we focus on the scatter of plastic behaviour. We statistically characterize the deformation process of micropillars using results from discrete dislocation dynamics (DDD) simulations. We then propose a stochastic microplasticity model which uses the extracted information to make statistical predictions regarding the micropillar stress-strain curves. This model aims to map the complex dynamics of interacting dislocations onto a stochastic processes involving the continuum variables of stress and strain. Therefore, it combines a classical continuum description of the elastic-plastic problem with a stochastic description of plastic flow. We compare the model predictions with the underlying DDD simulations and outline potential future applications of the same modelling approach.

## 1. Introduction

The miniaturisation of systems and devices creates the need to address the mechanical properties of materials on smaller and smaller scales. Figure 1 illustrates the differences between the stress-strain curve of a macroscopic Mo single crystal specimen and the stress-strain curves of micropillars of the same material. We observe that microplasticity differs from macroplasticity in two important aspects. The stress-strain curves of the micropillar samples exhibit strong fluctuations and on average the micropillar specimens are much stronger than the macroscopic sample.

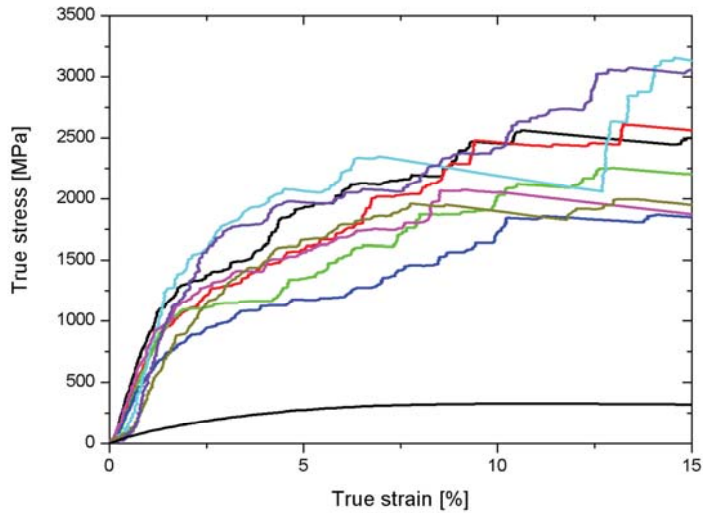

*Figure 1: Top: stress-strain curves of [100] oriented Mo micropillars, mean diameter d=0.3μm [4, 5]; Bottom: room-temperature stress-strain curve of macroscopic [100] oriented Mo single crystal [6].*

While a lot of effort has gone into understanding and modelling the size-dependent strength of small samples for both fcc and bcc materials (for recent reviews see [1-3]), the question of fluctuations in strength has been less investigated. It is clear that such an investigation ought to be based upon studying the dynamics of dislocations as the main carriers of plastic deformation in crystalline materials. The collective motion of dislocations occurs in an intrinsically jerky and intermittent manner. Even in macroscopic specimens, acoustic emission measurements reveal intermittent fluctuations of the energy release rate ("dislocation avalanches") whose magnitudes span over 6 decades in energy release [6], While, in macroscopic specimens, these fluctuations are not directly visible on the stress-strain curves, with decreasing sample size the intermittent avalanche-like dynamics of dislocations becomes directly visible in the form of stress drops or strain bursts punctuating the stress-strain curves.

A significant amount of papers have discussed the question how we should understand the term "plastic yielding" in small samples. Some studies argue that the yield stress corresponds to the occurrence of the first large avalanche [7] but, given that deformation bursts in microplasticity tend to follow power law distributions [8-11], it is not quite clear how to define a threshold for "large" avalanches in any meaningful manner. Maass et. al. suggest to associate yielding with the first observation of lattice rotations [8] but again, since *any* dislocation activity is associated with microscale lattice rotations, the problem of defining a threshold is not solved by this definition. Other studies refer to concepts drawn from statistical physics and envisage yielding as a depinning-like phase transition [8,9,12,13], though this idea has been recently questioned [14]. Despite the differences in interpretation, there is some consensus in the literature that the statistics of strain bursts in microplasticity can be meaningfully described by (truncated) power law distributions. In the present paper we refrain from entering controversies regarding interpretation - we simply determine the parameters of these distributions in a phenomenological manner to best reproduce stress strain curves obtained from DDD simulations. The same is done for the yield stress.

Theoretical approaches to micro-plasticity have mainly focused on the modelling of size effects, by including length scales into constitutive equations of plasticity [15-17] or more recently by formulating the dynamics of dislocations within a continuum framework [18,19]. Finally, an alternative approach to plasticity of micron-scale samples is provided by the method of discrete dislocation dynamics (DDD) simulation [20] which, while computationally demanding, provides complete information about stresses and strains on the dislocation scale and thus gives natural access to both size effects [21,22,23,24] and fluctuation phenomena [11,22].

Our proposition in the present manuscript is to generalise continuum theories by an appropriate stochastic description of the deformation process in order to include local variability. Following the ideas expressed in [25], we construct a stochastic model for the deformation behaviour based upon the statistical analysis of DDD simulations. The paper is organised as follows. Section 2 provides a description of the details of the 3D DDD simulations and illustrates the statistical analysis of the DDD data. Section 3 describes the stochastic model and evaluates its performance for different degrees of complexity of the statistical model. General conclusions are given in Section 4.

## 2. Statistical Analysis of DDD Simulations

For this work, we simulated strain-controlled tensile deformation of cubic samples with dimensions of 0.50 x 0.50 x 0.50 $\mu m^3$. The monocrystalline samples have face centered cubic (fcc) lattice structure, and their edges are oriented along the cubic axes of the crystal lattice. We impose a constant displacement rate on the upper sample surface, corresponding to an imposed strain rate (displacement velocity divided by specimen height) of 5000 $s^{-1}$. The bottom surface of the specimens remains fixed, and the side surfaces are free. The initial dislocation microstructures consist of 48 randomly distributed Frank-Read sources. On each slip system there are 4 sources of 0.22 μm length leading to an initial dislocation density of about $8 \times 10^{13} m^{-2}$. The material is assumed to have Young's modulus $E$ = 72.7 GPa (close to Al). Results of 22 different simulations with different initial source positions are shown in Figure 2.

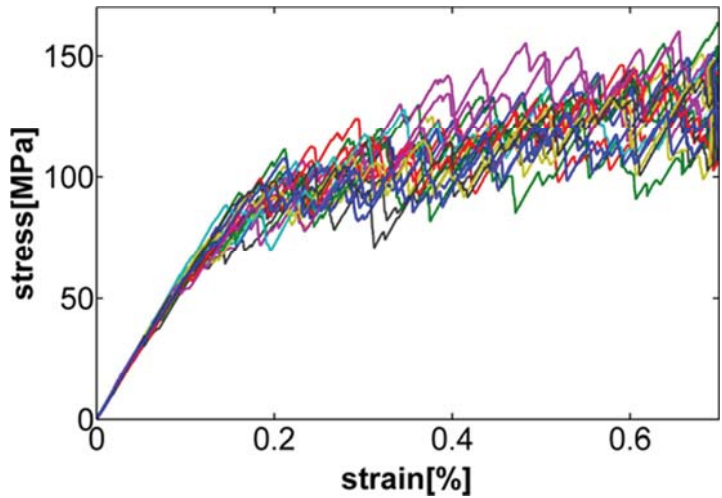

*Figure 2: Stochastic nature of plastic flow as illustrated by superimposing the stress-strain curves of 22 DDD simulations. For details see text.*

As the dislocation lines are randomly distributed in the sample at the beginning of the simulation, their interaction in the initial configuration leads to a certain amount of dislocation motion even in the absence of external action upon the system. The concomitant plastic relaxation strain, which may be either positive or negative, is offset in order to ensure that all DDD stress-strain curves start in the origin.

The simulated deformation curves can be divided into an initial, quasi-elastic part and a subsequent regime of plastic flow. Analysing the quasi-elastic part, we find that the slopes of the stress strain curves in this regime are less than the value of $E = 72.7$ GPa expected from the material's elastic constants. This can be readily understood by observing that, even before the yield stress is reached, dislocations undergo quasi-reversible (in Seeger's [26] terminology: "inversive") motion. Such motion reverses upon unloading in such a manner that the dislocation arrangement reaches its initial configuration and no permanent strain is produced. It is, however, thermodynamically irreversible since the loading-unloading curve encloses a finite area in the stress vs. strain plane. For illustration, we consider the sub-critical bowing out of a Frank-Read source of length $l$. Assuming for simplicity an isotropic line tension $T = Gb^2$, the critical configuration of the source (a semi-circle of a radius $l/2$ and area $\pi l^2/8$) is reached at a stress of $\tau=2Gb/l$. For an ensemble of Frank-Read sources of volume density $n$ (dislocation density $\rho=nl$) the corresponding "inversive" strain is given by

$$\varepsilon_{\text{inv}} = \frac{\pi n b l^2}{8M} = \frac{\pi}{8M}\rho b l, \qquad \frac{\varepsilon_{\text{inv}}}{\varepsilon_{\text{el}}} = \frac{\pi}{16}\rho l^2, \tag{1}$$

and the effective elastic modulus can be estimated as $E_{\text{eff}} = E/(1+ \varepsilon_{\text{inv}}/\varepsilon_{\text{el}})$. With $\rho l^2 \approx 1$ in our simulations we find an effective elastic modulus of about 60MPa, in good agreement with the simulated stress strain curves. Of course our estimate based on a consideration of single sources is an over-simplification since dislocation-dislocation interactions affect the stress response of the initial dislocation configuration. As a consequence, we find a not insignificant scatter of the initial slopes of the stress strain curves (cf. Section 3.3).

Above a critical stress the samples enter a plastic deformation regime where dislocation motion becomes irreversible (the initial configuration is not restored upon unloading). Again as a consequence of dislocation-dislocation interactions, the corresponding critical stresses fall significantly below the estimate of $\tau=2Gb/l$ for a single Frank-Read source. The ensuing plastic deformation regimes are characterized by strongly intermittent behaviour. Deformation proceeds as a discrete sequence of 'deformation events', so called avalanches [11] during which the plastic deformation rate increases significantly. During an avalanche the plastic strain rapidly increases and the stress decreases (3).

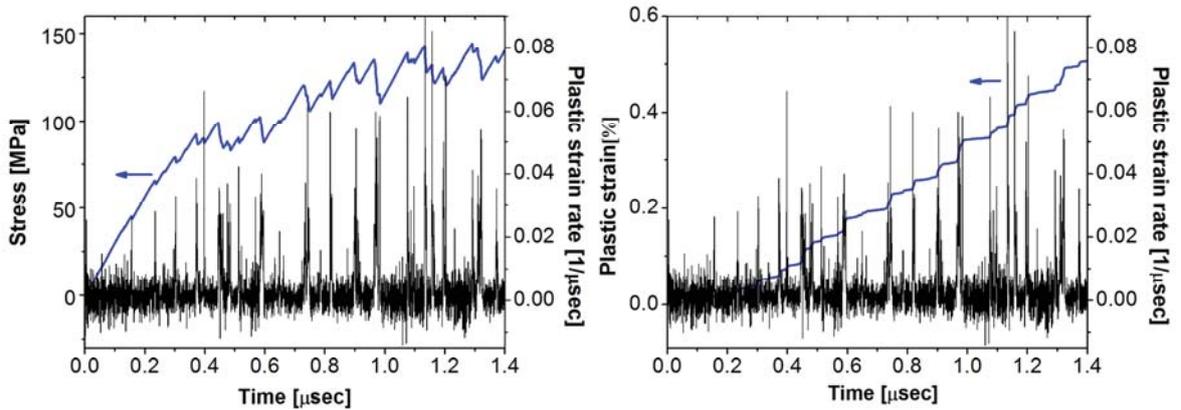

*Figure 3: Stress, strain rate and plastic strain vs. time signals in a DDD simulation of uniaxial compression; left, plastic strain vs. time and strain rate vs. time; right, stress vs. time and strain rate vs. time*

Figure 3 demonstrates the correlation between stress and plastic strain rate and the correlation between plastic strain and plastic strain rate, respectively. Clearly we are dealing with two different regimes of dynamic behavior: the avalanches/strain bursts and the intervals in between. The first step towards the statistical characterization of the stress-strain curves is therefore to separate our time records into active and inactive parts. The active parts are the time intervals during which avalanches occur. The inactive parts are the intervals between the avalanches. Firstly, we smooth all the time records by an averaging process of adjacent points. This serves to eliminate the rapid oscillations which stem from the discrete timestepping of the DDD code and are thus numerical artefacts. We note that an analogous procedure may be needed in analysing experimental data where comparable oscillations may arise from the mechanical action and electronic control of the microdeformation rig [27].

Excluding the initial elastic loading part we separate the DDD simulation records into "active" and "inactive" time intervals by imposing a threshold value on the plastic strain rate. By choosing this threshold to equal the imposed strain rate, the former correspond to decreasing and the latter to increasing parts of the stress vs. total strain curve. Thus, a strain burst or dislocation avalanche is, in the present analysis, simply defined as a time interval over which the plastic strain rate exceeds the imposed value. Subsequently, we determine the changes in stress and in plastic strain which occur during the active and inactive time intervals. The resulting records can be statistically characterized in terms of probability distributions of the respective variables. In order to determine these probability distributions we use rank ordering statistics [28].

We make an important simplification: Figure 3 (right) indicates that, during an inactive time interval, the plastic strain is approximately constant. Similarly, Figure 3 (left) shows that the duration of an "active" time interval is much less than the duration of an "inactive" one. Accordingly, we assume the plastic strain change during an inactive interval and the total strain change during an active interval to be equal to zero. This implies that the entire sequence can be characterized in terms of two sequences of variables only: The plastic strain increments $\Delta\varepsilon_a$ during the "active" time intervals, and the stress increments $\Delta\sigma_i$ during the "inactive" intervals. Since the deformation is assumed to be instantaneous during the "active" and purely elastic during the inactive intervals, $\Delta\varepsilon_a$ also corresponds to a stress drop $\Delta\sigma_a \approx -E\Delta\varepsilon_a$ and $\Delta\sigma_i$ to a total strain increment $\Delta\sigma_i/E$ where $E$ is the Young's modulus of the simulated samples.

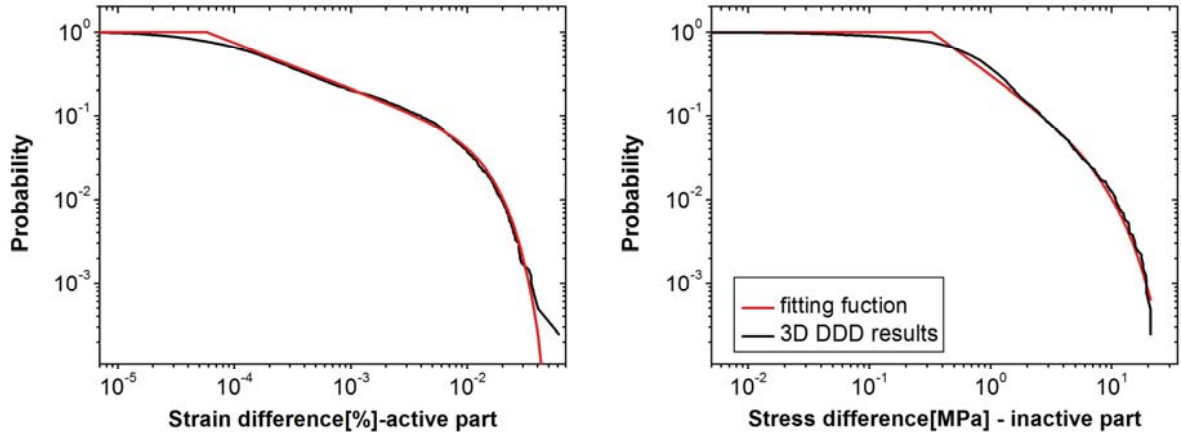

*Figure 4: Left, rank-ordered distribution of plastic strain increments during strain bursts ("active" time intervals); right, distribution of stress increments during inactive time intervals, determined from 22 DDD simulations of uniaxial compression as described in the text. Black line corresponds to simulation data; red line corresponds to fitting function*

We collected data from multiple simulations and evaluated by rank ordering the probability distributions of the plastic strain differences $\Delta\varepsilon_a^n$ and the stress changes $\Delta\sigma_i^n$. To each value of one of these variables we assign its position ($k$) in a list ordered by decreasing magnitude. The probability $p(X_{(k)})$ to find a value of the random variable $X$ less than $X_{(k)}$ is then estimated as

$$p(X_{(k)}) \approx \frac{k}{M+1} \quad (2)$$

where $M$ is the total number of entries in the list. The resulting probabilities $p(\Delta\varepsilon_a)$ and $p(\Delta\sigma_i)$ are shown in double logarithmic plots in Figure 4. The black curves present the data and the red curves the respective fitting functions. Comparing the two graphs we note that there is a remarkable degree of similarity between the statistics of both stress and plastic strain increments.

It is well established that plastic strain increments produced by slip avalanches can be characterized by truncated power law distributions [8,10]. In our case both $p(\Delta\varepsilon_a)$ and $p(\Delta\sigma_i)$ seem to be well described by truncated power laws,

$$p(\Delta\varepsilon_a) = \left(\frac{\Delta\varepsilon_a}{\Delta\varepsilon_a^{min}}\right)^{-a} \cdot exp\left(-\left[\frac{\Delta\varepsilon_a}{\Delta\varepsilon_a^{max}}\right]^b\right), \Delta\varepsilon_a > \Delta\varepsilon_a^{min}; \quad p(\Delta\varepsilon_a) = 1 \;\; otherwise;$$

$$p(\Delta\sigma_i) = \left(\frac{\Delta\sigma_i}{\Delta\sigma_i^{min}}\right)^{-a} \cdot exp\left(-\left[\frac{\Delta\sigma_i}{\Delta\sigma_i^{max}}\right]^b\right), \Delta\sigma_i > \Delta\sigma_i^{min}; \quad p(\Delta\sigma_i) = 1 \;\; otherwise. \quad (3)$$

The fitting parameters (red curves) are given in Table 1.

|  | *min* | *max* | *a* | *b* |
|---|---|---|---|---|
| $\Delta\varepsilon_a$ | $5{,}84 \cdot 10^{-7}$ | $1{,}70 \cdot 10^{-4}$ | 0,55 | 1,8 |
| $\Delta\sigma_i$ | $3.3 \cdot 10^5$ | $1 \cdot 10^7$ | 1.05 | 1.5 |

*Table 1: Fitting parameters for the probability distributions in Figure 4*

## 3. Stochastic Model

### 3.1 Naïve model: Uncorrelated avalanche sequence

The aim of a stochastic microplasticity model is to map the complex dynamics of interacting dislocations onto stochastic processes involving the continuum variables of stress and strain. Using statistical information extracted from DDD, our stochastic model is constructed to reproduce the essential statistical features of the deformation processes in small volumes of a material. The simplest conceivable model is to assume that the variables characterizing the "active" and "inactive" intervals which alternate above the yield stress each represent stationary, uncorrelated stochastic point processes, with probability distributions given by Eq. (3).

In a strain controlled tension stochastic simulation the stress strain curve then consists of an initial elastic part up to a yield stress $\sigma = 60$ MPa which is chosen to optimally match, on average, the DDD simulations. Afterwards, the deformation curves consists of alternating segments: During an active interval, the total strain remains constant, the plastic strain increases by an amount $\Delta\varepsilon_a$ randomly drawn from the distribution $p(\Delta\varepsilon_a)$, equation 2, and the stress decreases by $E\Delta\varepsilon_a$. During the subsequent inactive interval, stress increases by an amount $\Delta\sigma_i$ randomly drawn from the distribution $p(\Delta\sigma_i)$, equation 3. The plastic strain remains constant and the total strain increases by $\Delta\sigma_i/E$. This process is repeated until the desired end strain is reached. No correlation is assumed between $\Delta\sigma_i$ and $\Delta\varepsilon_a$, or between the sequential values of either of these variables. A stress-strain curve simulated using this simple algorithm is shown in Figure 5. (We note that the model has close similarities to the SUDTS algorithm proposed by Kugiumtzis and Aifantis for constructing random surrogates to stress-strain curves in macroscopically jerky plastic flow [29].)

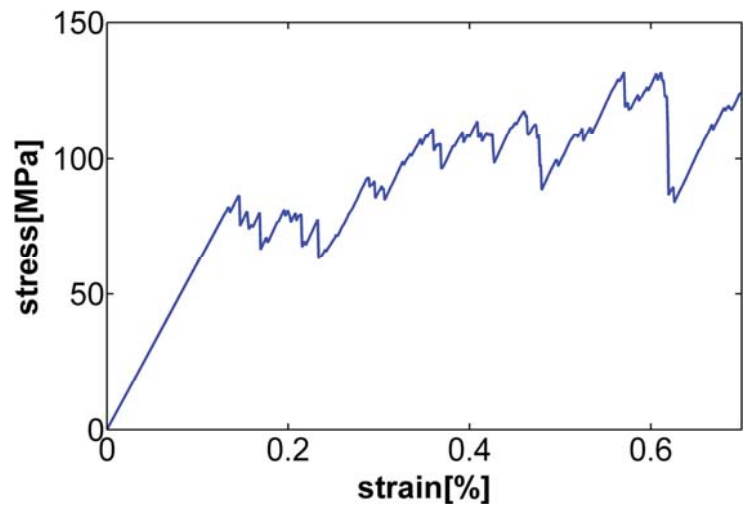

*Figure 5: A stress-Strain curve calculated from the uncorrelated stochastic model*

To quantitatively compare results obtained from the stochastic model with those from 3D DDD simulation we consider the mean and the standard deviation of stress calculated as functions of total strain for ensembles of both DDD and of stochastic simulations.

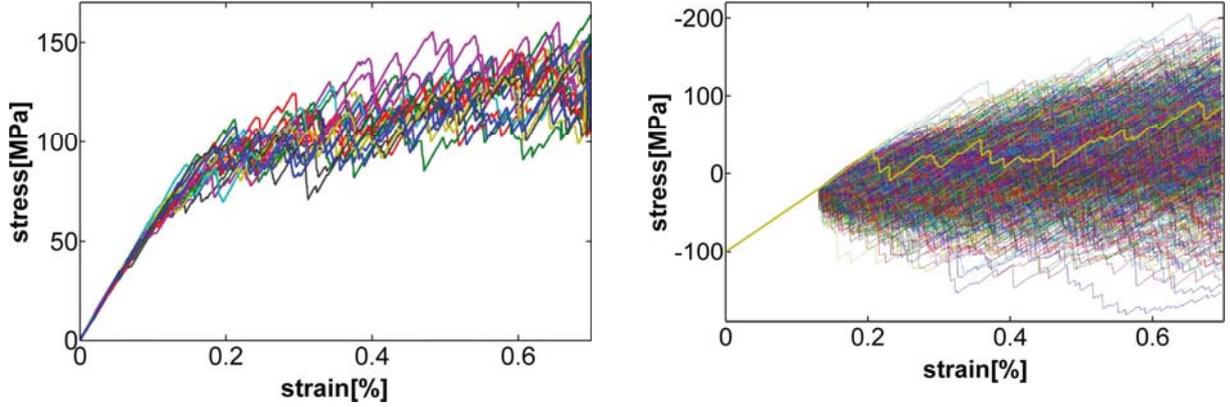

*Figure 6: Comparison of stochastic simulation results (left) with the stress strain curves of 22 DDD simulations (right, see also Figure 2)*

Figure 6 illustrates the results of 1000 different stochastic simulations as compared with the results of the 22 3D DDD simulations.

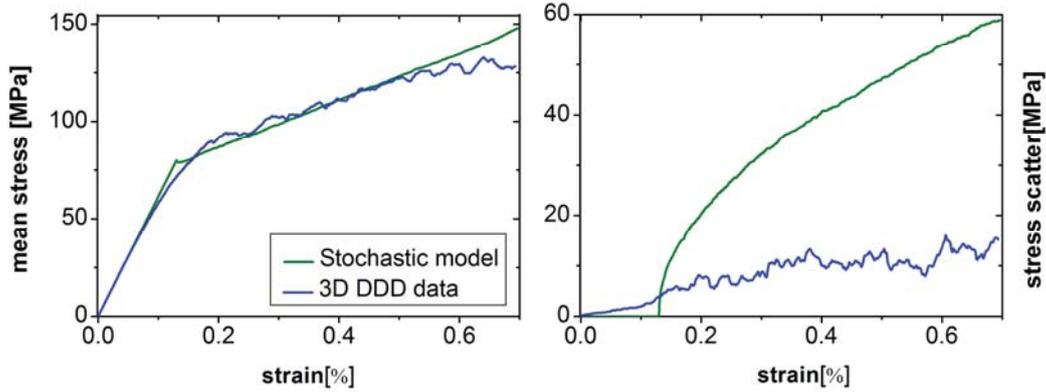

*Figure 7: Right, average stress of DDD simulations (blue line) and stochastic simulations (green line); Left, stress standard deviation of DDD simulations (blue line) and stochastic simulations (green line)*

*Figure*7 compares the mean values and standard deviations of stresses obtained from the DDD and stochastic simulation ensembles. The fact that the curves for the stochastic ensemble are much smoother is simply due to the much larger number of simulations (1000 stochastic vs. 22 DDD simulations). We first focus on the mean stress-strain curves. The average stress and plastic strain reached after *n* active-inactive cycles are given by

$$\langle \sigma \rangle = \sigma_y + n(\langle \Delta\sigma_i \rangle - E\langle \Delta\varepsilon_a \rangle), \quad \langle \varepsilon \rangle = n\langle \Delta\varepsilon_a \rangle \qquad (4)$$

It follows that the ensemble-averaged hardening coefficient is given by

$$\theta = \frac{d\langle \sigma \rangle}{d\langle \varepsilon \rangle} = \frac{\langle \Delta\sigma_i \rangle}{\langle \Delta\varepsilon_a \rangle} - E \qquad (5)$$

The calculation shows that our stochastic model should produce, on average, linear hardening above yield. As can be seen from Figure 7, this result is in line with the stochastic simulation data. It is also seen that the averaged stress strain curves from the stochastic model and the DDD simulations are in reasonable agreement.

This is not true for the statistical scatter of flow stresses, which is very significantly over-estimated by the present, simplistic model. The long-time behavior of the stress in this model can be envisaged as a random walk superimposed on a linear trend. For small strains, the parabolic growth of the stress scatter (diffusion-like behavior of the random walker, see Figure 7, right) predominates to such an extent that, in a non-negligible fraction of all realizations of the stochastic model, negative flow stresses are reached (Figure 6). Of course this is completely unphysical. We may thus conclude that the assumption of a sequence of active and inactive periods with uncorrelated stress and strain increments does not adequately represent the DDD simulations.

### 3.2 Correlated Stochastic Model

In order to improve the stochastic model we more closely examine the assumption that the stress changes during "active" and "inactive" parts are uncorrelated random variables. Consider a single DDD simulation as shown in Figure 8 (right). We plot the stress changes during the active and inactive time intervals versus the index $n$ which indicates the position of an interval in the respective record (i.e., the first stress drop and the subsequent stress increase both have index $n=1$, the second stress drop and its subsequent stress increase have index $n=2$, etc.). We can distinguish two regimes: At the onset of deformation, we see pronounced stress increases but only very small stress drops. This corresponds to the microplastic regime before yield, where we essentially see elastic loading punctuated by few and small stress drops. Above the yield stress, the picture changes: In this regime, we see that large stress drops tend to be followed by large stress increases – hence, there is a significant correlation between the stress drop and stress increase corresponding to a given "active-inactive-cycle".

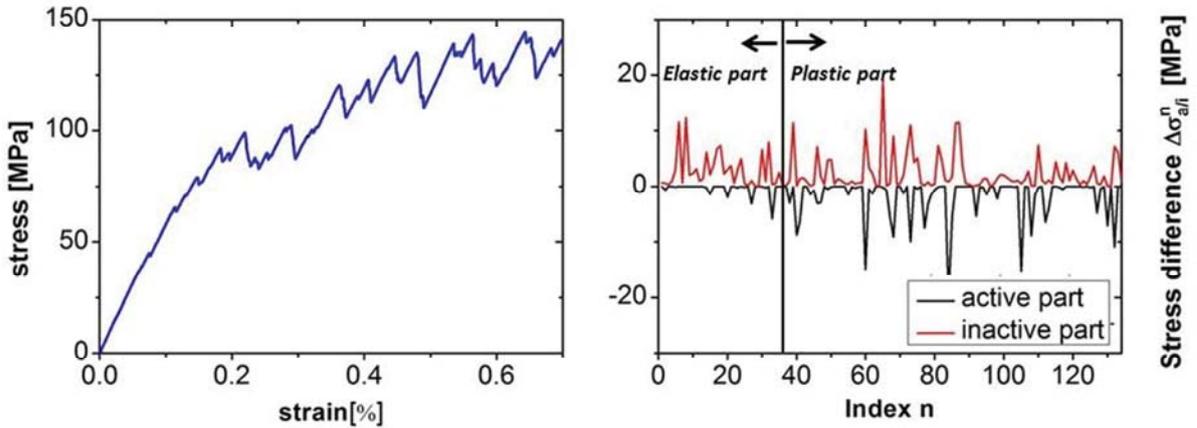

*Figure 8: right, black line: stress difference during the active part of an active-inactive cycle vs. cycle number n; red line: stress difference during inactive part vs. cycle number n.*

These observations lead to the definition of an improved stochastic model where, in the plastic regime, the variables characterizing a stress drop/ strain burst and the subsequent elastic stress increase within an active-inactive cycle possess some degree of correlation. The elastic part of the stress-strain curve, on the other hand, is still envisaged as a straight line up to the yield point. As previously, the yield stress has a fixed value which is chosen to best match the DDD simulations.

In constructing correlated random variables $\Delta\varepsilon_a^n$ and $\Delta\sigma_i^n$, we face the problem that these variables are not identically distributed. We therefore perform an intermediate step where we construct two correlated Gaussian variables $L_{1,2}$ with Pearson correlation coefficient $q$ and then convert these to uniformly distributed variables $Y_{1,2}$ using the probability integral transform [30],

$$Y_{1,2} = \Phi(L_{1,2}) = \frac{1}{\sqrt{2}} \int_{-\infty}^{L_{1,2}} e^{\frac{-t^2}{2}} dt = \frac{1}{2}\left[1 + \text{erf}\left(\frac{L_{1,2}}{\sqrt{2}}\right)\right]. \tag{6}$$

From these we obtain correlated values of $\Delta\varepsilon_a$ and $\Delta\sigma_i$ by setting $p(\Delta\varepsilon_a) = Y_1$ and $(\Delta\sigma_i) = Y_2$.

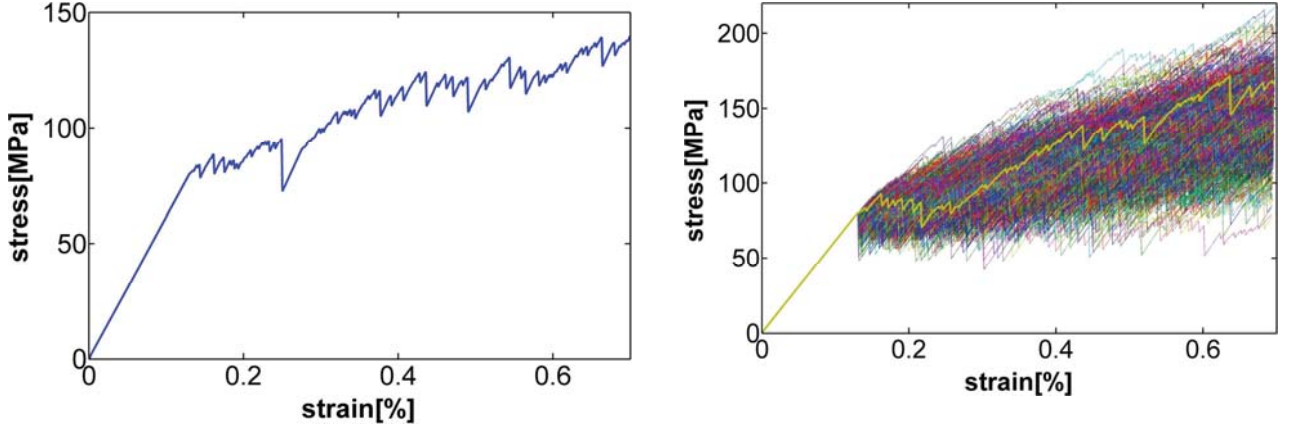

*Figure 9: Left, stress strain curve calculated from the correlated stochastic model for correlation factor equal to 1; right, 1000 stress strain curves from correlated stochastic simulations with correlation factor q=1, the elastic part coincides for all the simulations*

The stress strain curve of a stochastic simulation with $q = 1$ is shown in Figure 9 and exhibits an interesting shape. The stress decreases during the active part and the stress increases during the inactive part are now always of the same order of magnitude – basically, stress drops due to the plastic strain increment are mostly reversed during the subsequent quasi-elastic stress increase. This behaviour is expected if we consider that, in the fully developed plastic regime, the dislocation microstructures before and after a strain burst avalanche are statistically to a large degree equivalent – hence, we expect the strain burst initiation stresses to be not too different. The same effect also prevents the stress-strain curves in different simulations from drifting too far apart and prevents the simulations from straying into the unphysical regime of negative flow stresses (compare Figure 9 (left) with Figure 6).

To investigate the performance of the modified model we again compare the mean and the standard deviation of stress as functions of strain as obtained from ensembles of DDD simulations and of correlated stochastic simulations, now for different correlation factors.

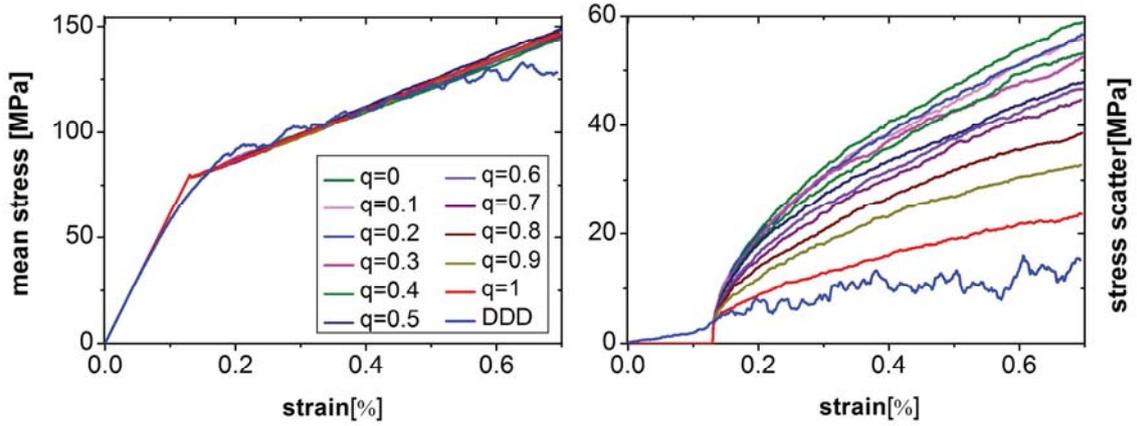

*Figure 10:Left, average stress of DDD simulations (blue line) and correlated stochastic simulations (coloured lines according to the correlation factor); right, average stress of DDD simulations (blue line) and correlated stochastic simulations (coloured lines according to the correlation factor)*

Figure 10 (right) shows the mean stress as a function of strain. Here there is no influence of the correlation factor. This is to be expected since equations (4) and (5) do not depend on the presence or absence of correlations. However, correlations have a significant impact on the scatter of the stress-strain curves, as seen in Figure 1 (left). As the correlation between stress drops and stress increases becomes more pronounced, the scatter of the stress-strain curves obtained from the stochastic model decreases and approaches the scatter of the DDD simulations. In other words, as the correlation increases between the active and inactive intervals the model becomes more reliable in reproducing the fluctuations around the mean stress level. Still, our model is amenable to improvements since the assumption of fully deterministic behaviour up to a uniform yield point is clearly unrealistic – in the DDD simulations we see a gradual, rather than an abrupt onset of scatter in the stress strain curves. In the following section we explore the possibility of including statistical scatter in dislocation behaviour *before* yield.

### 3.3 Stochastic model with fluctuations in dislocation behavior before yield

As discussed above in Section 2, the initial slope of the stress-strain curves is influenced by "inversive" dislocation motions. These motions do not lead to slip avalanches, but can rather be envisaged as a quasi-reversible polarization of the initial dislocation configuration that occurs once a stress is applied. Owing to the randomness of the initial configuration, the strain produced by such motions – and accordingly the effective elastic modulus – exhibits statistical scatter. This is shown in Figure 11 which shows the distribution of effective elastic moduli $E$ in the DDD simulations, defined as the ratio between axial stress and total axial strain at an axial strain of 0.001 which is below the strain where large-scale plastic yielding occurs (see Figure 6, left). We see that the average value $E \sim 58$ MPa agrees well with the estimate provided above in Section 2. At the same time, there is some statistical scatter around this value.

We now construct a simple stochastic model as follows: We draw an effective elastic modulus from the distribution shown in Figure 11 which we approximate by a Weibull distribution

$$P(E) = \exp\left(-\left(\frac{E-E0}{E1}\right)^{3.5}\right) \qquad (7)$$

where $E_0$ = 52 GPa and $E_1$ = 5.5 GPa. We then apply a purely elastic deformation up to a yield strain of 0.13 chosen to match the average behaviour of the DDD curves. Above this yield strain we continue with a simulation as in Section 3.2. Note that this procedure has the consequence that yield stresses in our simulation are Weibull distributed with an exponent of about 3.5, as previously proposed by other authors for interpreting micropillar experiments and simulations (see e.g. [31,32]).

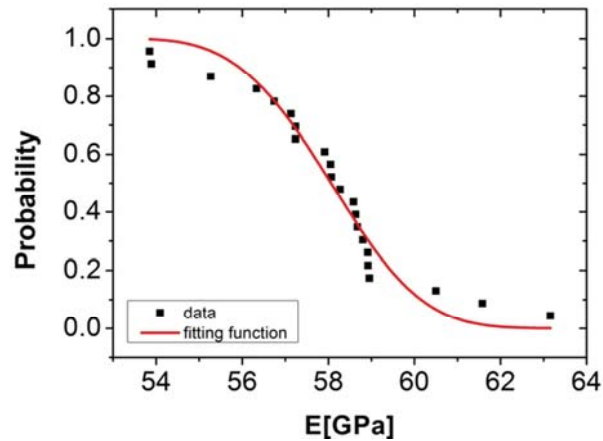

*Figure 11: Rank-ordered distribution of effective elastic moduli in the DDD simulations. Data points: simulation data; red line corresponds to fit assuming a Weibull distribution.*

Comparing the mean and scatter of the flow stress vs strain curves obtained from this model with those obtained from DDD demonstrates that the behaviour both before and after yield is now well represented. In particular the gradual increase of the scatter before yield is now well represented (Figure 12).

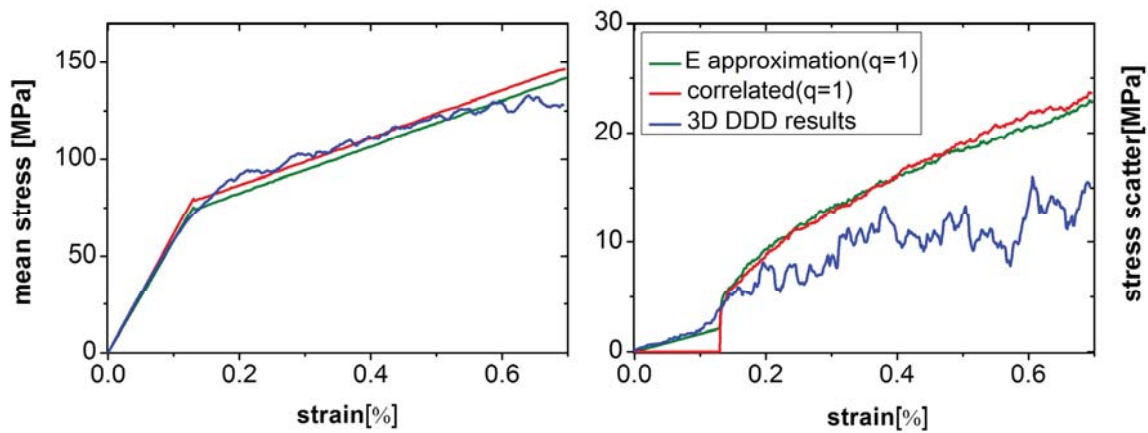

*Figure 12: Left: average stress; Right: stress standard deviation; Blue line: DDD simulations results; Red line: correlated stochastic simulations using correlation factor q=1; Green line: correlated stochastic simulations using correlation factor q=1 and statistically distributed effective elastic moduli.*

## 4. Summary and conclusions

We have explored some simple models describing stress-strain curves in microplastic deformation as stochastic processes consisting of quasi-elastic stress increases and sudden plastic strain increments/stress drops. Such models can provide a reasonably good approximation of the behavior found in DDD simulations, provided that a strong correlation is assumed between a stress drop/plastic strain burst and the following stress increase. However, even in this case a certain over-estimation of the fluctuations persists. This indicates that longer-term correlations are present in the dislocation dynamics which encompass multiple slip events and delimit fluctuations.

A strong limitation of the model consists in the assumption that the stress-strain curves can be modelled in terms of stationary stochastic processes. In materials undergoing significant strain hardening, this approximation can in general not be sustained since the dislocation microstructure may undergo significant changes in the course of strain hardening. In the DDD simulations considered here, however, no very significant increase in dislocation density is observed, such that over the limited strain ranges attainable in the DDD simulations the assumption of stationarity after yield may be considered acceptable.

**Acknowledgement:** We acknowledge support by Deutsche Forschungsgemeinschaft DFG-FOR1650, grant WE3544/5-1, and by EPSRC under grant no. Ep/J003387/1.

## References:

1.  Greer, J.R. and J.T.M. De Hosson, *Plasticity in small-sized metallic systems: Intrinsic versus extrinsic size effect*. Progress in Materials Science, 2011, **56,** 654-724.
2.  Kraft, O., P.A. Gruber, R. Mönig and D. Weygand, *Plasticity in confined dimensions*. Annual Review of Materials Research, 2010, **40**, 293-317.
3.  Uchic, M.D., P.A. Shade, and D.M. Dimiduk, *Plasticity of Micrometer-Scale Single Crystals in Compression.* Annual Review of Materials Research, 2009, **39**, 361-386..
4.  Schneider, A.S., et al., *Effect of orientation and loading rate on compression behavior of small-scale Mo pillars*. Materials Science & Engineering A, 2009, **508**: 241-246.
5.  Hollang, L., D. Brunner, and A. Seeger, *Work hardening and flow stress of ultrapure molybdenum single crystals*. Materials Science & Engineering A, 2001, **319**: 233-236.
6.  Miguel, M.C., A. Vespignani, S. Zapperi, J. Weiss, and J.R. Grasso, *Intermittent dislocation flow in viscoplastic deformation*, Nature, 2001, **410**, 667-671.
7.  Dimiduk, D.M., M.D. Uchic, and T.A. Parthasarathy, *Size-affected single-slip behavior of pure nickel microcrystals.* Acta Materialia, 2005, **53,** 4065-4077.
8.  Nikitas, N. and M. Zaiser, *Slip avalanches in crystal plasticity: scaling of the avalanche cutoff.* J. Stat. Mech: Theory and Experiment, 2007, P04013.
9.  Zaiser, M. and P. Moretti, *Fluctuation phenomena in crystal plasticity - a continuum model.* J. Stat. Mech: Theory and Experiment, 2005, P08004.
10. Csikor, F.F., et al., *Dislocation avalanches, strain bursts, and the problem of plastic forming at the micrometer scale.* Science, 2007. **318**, 251-254.
11. Maaß, R., et al., *Smaller is stronger: The effect of strain hardening.* Acta Materialia, 2009. **57**, 5996-6005.
12. Zaiser, M., *Scale invariance in plastic flow of crystalline solids*. Adv. Phys., 2006, **54**, 185-245.
13. Chan, P. Y. et. al., *Plasticity and dislocation dynamics in a phase field crystal model*. Phys. Rev. Letters, 2010, **105**, 015502.
14. Ispánovity, P. D. et. al., *Avalanches in 2D Dislocation Systems: Plastic Yielding Is Not Depinning.* Physical review letters, 2014, **112**, 235501.
15. Aifantis, E., *On the microstructural origin of certain inelastic models*. 1984. p. 131-149.
16. Fleck, N. and J. Hutchinson, *Strain gradient plasticity.* Advances in applied mechanics, 1997. **33**: p. 295-361.
17. Gurtin, M.E., *On the plasticity of single crystals: free energy, microforces, plastic-strain gradients.* Journal of the Mechanics and Physics of Solids, 2000. **48**(5): p. 989-1036.

18. Zaiser, M., et al., *Modelling size effects using 3D density-based dislocation dynamics.* Philosophical Magazine, 2007. **87**(8-9): p. 1283-1306.
19. Hochrainer, T., S. Sandfeld, M. Zaiser and P. Gumbsch, *Continuum dislocation dynamics: towards a physical theory of crystal plasticity.* Journal of the Mechanics and Physics of Solids, 2014, **63**, 167-178.
20. Weygand, D., et al., *Aspects of boundary-value problem solutions with three-dimensional dislocation dynamics*. 2002.
21. Weygand, D. et al., *Three-dimensional dislocation dynamics simulation of the influence of sample size on the stress–strain behavior of fcc single-crystalline pillars.* Mater. Sci. Engng. A, 2008, **483**, 188-190.
22. Senger, J. et al., *Aspect ratio and stochastic effects in the plasticity of uniformly loaded micrometer-sized specimens.* Acta Mater., 2011, **59**, 2937-2947.
23  Rao SI, Dimiduk DM, Parthasarathy TA, Uchic MD, Tang M, Woodward C. 2008. *Athermal mechanisms of size-dependent crystal flow gleaned from three-dimensional discrete dislocation simulations.* Acta Mater. 56:3245–59
24  El-Awady JA, Wen M, Ghoniem NM. 2009. *The role of the weakest-link mechanism in controlling the plasticity of micropillars.* J. Mech. Phys. Solids. 57:32–50
25. Zaiser, M., *Statistical aspects of microplasticity: experiments, discrete dislocation simulations and stochastic continuum models*, J. Mech. Behavior Mater, 2013, **22**, 89-100.
26. Zaiser, M. and A. Seeger, *Long-range internal stresses, dislocation patterning and work hardening in crystal plasticity*. In: F.N.R. Nabarro and M.S. Duesbery (Eds.), Dislocations in Solids Vol. 11, Elsevier 2002, pp. 1-100.
27. Zaiser, M., J. Schwerdtfeger, A.S. Schneider, C.P. Frick, B.G. Clark, P. Gruber and E. Arzt, *Strain bursts in plastically deforming molybdenum micro-and nanopillars.* Philosophical Magazine, 2008, **88**, 3861-3874.
28. David, H.A. and H.N. Nagaraja, *Order statistics*. 1970: Wiley Online Library.
29. Kugiumtzis, D., and Aifantis, E. C., *Statistical analysis for long term correlations in the stress time series of jerky flow.* J. Mech. Behavior Mater., 2004, **15**, 135-148.
30. Aickelin, U., *The Oxford dictionary of statistical terms.* Journal of the Operational Research Society, 2004. **55**: 1014-1014.
31. Ispánovity, P. D., Hegyi, Á., Groma, I., Györgyi, G., Ratter, K., and Weygand, D. (2013). *Average yielding and weakest link statistics in micron-scale plasticity*. Acta Mater., 2013, 61, 6234-6245.
32. Rinaldi A, Peralta P, Friesen C, Sieradzki K. Acta Mater., 2008, **56**, 511–517